\documentclass[aps,showpacs,floatfix,prl,reprint]{revtex4-1}

\usepackage{graphicx,bm}
\usepackage{subfigure}
\usepackage{amsmath}
\usepackage{amssymb}
\newcommand{\ket}[1]{|#1\rangle}
\newcommand{\bra}[1]{\langle #1|}
\newcommand{\scal}[2]{\langle #1|#2\rangle}
\newcommand{\brkt}[3]{\bra{#1} #2 \ket{#3}}
\newcommand{\moy}[1]{\bigl\langle #1 \bigr\rangle}
\newcommand{\cc}[1]{\hat{c}^\dagger_{#1}}
\newcommand{\ca}[1]{\hat{c}^{\,}_{#1}}
\renewcommand{\H}{\hat{H}}
\newcommand{\T}{\hat{T}}
\newcommand{\Os}{\hat{O}_s}
\newcommand{\ws}{\omega_s}
\newcommand{\R}{\mathcal{R}}
\newcommand{\bbE}{\mathbb{E}}
\renewcommand{\d}{\mathrm{d}}
\newcommand{\proptauinf}{\underset{\tau\to\infty}{\propto}}
\newcommand{\eqtauinf}{\underset{\tau\to\infty}{=}}
\newcommand{\e}{\mathrm{e}}

\begin{document}

\title{A Constrained-Path Quantum Monte-Carlo Approach \\ for the Nuclear Shell Model}
\author{J.~Bonnard}
\author{O.~Juillet}
\affiliation{Laboratoire LPC Caen, ENSICAEN, Universit{\'e} de Caen, CNRS/IN2P3,
6 Boulevard Mar{\'e}chal Juin, 14050 Caen Cedex, France}

\begin{abstract}
A new Quantum Monte-Carlo (QMC) approach is proposed to investigate low-lying states of nuclei within the shell model.
The formalism relies on a variational symmetry-restored wave-function to guide the underlying Brownian motion.
Sign/phase problems that usually plague QMC fermionic simulations are controlled by constraining stochastic paths through a fixed-node like approximation.
Exploratory results in the $sd$ and $pf$ valence spaces with realistic effective interactions are presented.
They prove the ability of the scheme to yield nearly exact yrast spectroscopies for both even- and odd-mass nuclei.
\end{abstract}
\date{\today}
\maketitle

The shell model with configuration interaction \cite{Caurier_SM} is one of the most powerful frameworks for nuclear structure calculations.
In this picture, the nucleons beyond an inert magic core are confined in an active shell and interact through an effective two-body residual potential.
The wave-functions of the nucleus are then determined by diagonalization of the Hamiltonian in the set of all the possible configurations and used to calculate physical observables.
Unfortunately, the exponential scaling of the many-body space with the number of valence nucleons or the size of the single-particle basis puts strong restrictions on the applicability of the shell model.

Quantum Monte-Carlo (QMC) methods potentially offer attractive alternatives to such limitations.
The many-body problem is indeed reduced to a set of stochastic one-body problems, numerically tractable, describing independent particles that randomly walk in fluctuating external fields.
In this way, exact correlated wave-functions are reconstructed by averaging independent-particle states called \emph{walkers}.
To date, the Shell Model Monte-Carlo method (SMMC) is the main application of a QMC approach for the nuclear shell model \cite{Koonin_SMMC}.
It relies on the standard sampling by auxiliary-fields of the path-integrals resulting from the Hubbard-Stratonovich transformation of the imaginary-time propagator.
With schematic residual interactions, the SMMC method accurately reproduces the properties of nuclei at zero and finite temperature \cite{Ozen_SMMC}.
However, with realistic effective interactions, the approach is plagued by the so-called fermion sign/phase problem which signature is an exponentially vanishing signal-to-noise ratio as the temperature is decreased.
To overcome this pathology, a family of modified Hamiltonians is usually sampled where interactions leading to the problem are artificially reduced.
The SMMC results have then to be extrapolated to recover exact observables \cite{Alhassid_SMMC_SP}.
Moreover, the SMMC method may not be used to obtain a detailed spectroscopy of nuclei.
In contrast, low-lying states can be reconstructed as long as the stochastic process is used only to generate a subspace in which the Hamiltonian is diagonalized. This so-called Monte-Carlo Shell Model technique \cite{Honma_MCSM} has been successfully applied to very large configuration spaces and provides variational estimates of nuclei properties at low energy \cite{Otsuka_MCSM}.

In this Letter, we present a new QMC scheme to investigate yrast states in the framework of the shell model.
The sign/phase problem is managed \textit{via} a constraint on the Brownian motion involving a trial wave-function with restored symmetries in the spirit of fixed-node \textit{ab initio} calculations \cite{Pieper_GFMC} where the random walk takes place in the position space.

Let us first consider a generic two-body Hamiltonian $\H$ cast in a quadratic form of one-body operators $\T$ and $\{\Os\}$:
\begin{equation}
   \H =\T -\sum_s \ws\Os^2 ,\; \T =\sum_{i,j} T_{ij} \cc{i}\ca{j} ,\; \Os =\sum_{i,j} (O_s)_{ij} \cc{i}\ca{j},
\end{equation}
where $\cc{i}$ ($\ca{i}$) is the creation (annihilation) operator of a fermion in a single-particle state $\ket{i}$ of an orthonormal basis.
This decomposition can be achieved by performing a Pandya transformation, as detailed in Ref. \cite{Koonin_SMMC}.

All QMC approaches use a stochastic representation of the imaginary-time propagator $\exp(-\tau\H)$ to project an initial wave-function $\ket{\Phi_0}$, supposed here to be a Slater determinant, onto the ground state $\ket{\Psi_G}$: $\lim_{\tau\to\infty} \exp(-\tau\H) \ket{\Phi_0} \propto \ket{\Psi_G}$.
In order to improve the efficiency, Zhang and Krakauer \cite{Zhang_Phaseless} have proposed to borrow the idea of importance sampling by generating walkers $\ket{\Phi}$ according to their overlap with a trial state $\ket{\Psi_T}$ not orthogonal to the ground state.
By directly including the complex importance function $\scal{\Psi_T}{\Phi}$ in the Brownian motion of walkers, one is led to reformulate the exact state at a given imaginary time $\tau$ according to
\begin{equation} \label{Ph_scheme}
   \exp \Bigl(-\tau\H\Bigr)\ket{\Phi_0} = \bbE \left[ \Pi_\tau \dfrac{\ket{\Phi_\tau}}{\scal{\Psi_T}{\Phi_\tau}} \right],
\end{equation}
where $\bbE[\cdot]$ denotes the average of a stochastic process.
The weight $\Pi_\tau$ is a Boltzmann-type exponential factor, and the evolution of the single-particle states $\{\ket{\phi_n}\}$ of the Slater determinant $\ket{\Phi}$ are driven by the following Ito stochastic differential equations:
\begin{equation} \label{Ph_scheme_2}
  \left\lbrace
  \begin{aligned}
     & \Pi_\tau =\scal{\Psi_T}{\Phi_0} \exp \left[- \int_0^\tau \d\tau' \moy{\H}_{\Psi_T,\!\Phi_{\tau'}} \right] \\
     & \d\ket{\phi_n} = \Bigl[-\d\tau \Bigl( T- \sum_s\ws\bigl[O_s^2+ \moy{\Os}_{\Psi_T,\!\Phi} O_s\bigr] \Bigr) \biggr. \\
     & \qquad\biggl.\qquad\qquad\qquad\qquad + \sum_s \sqrt{2\ws}\d W_s O_s \Bigr] \ket{\phi_n}.
   \end{aligned}
   \right. 
\end{equation}
Here, we introduced the so-called local estimate $\moy{\hat{A}}_{\Psi_T,\!\Phi}=\brkt{\Psi_T}{\hat{A}}{\Phi}/\scal{\Psi_T}{\Phi}$ of any observable $\hat{A}$.
The $\{W_s\}$ refer to independent Wiener processes with vanishing ensemble averages and the following multiplication table for their infinitesimal increments \cite{Gardiner}
\begin{equation}
   \forall\,s,s' \: : \: \bbE[\d W_s] = 0 \quad ; \quad \d W_s\d W_{s'} = \delta_{ss'}\d\tau \;.
\end{equation}
The stochastic dynamics defined by \eqref{Ph_scheme_2} differs from the underlying one of the standard SMMC method by the local estimate of the $\Os$ operators in the drift term.
Therefore, through this additional term, $\ket{\Psi_T}$ guides the random walk towards a region of the over-complete manifold of Slater determinants where the contribution to $\ket{\Psi_G}$ is expected to be large.

From a general point of view, any QMC sampling is plagued by sign/phase problems when the phase of the overlap $\scal{\Psi_G}{\Phi}$ varies during the Brownian motion.
In that case, a population can yield a mean overlap $\bbE[\scal{\Psi_G}{\Phi}]=0$, the contributions of these realizations cancelling each others.
Hence, such trajectories are useless for reconstructing the ground-state wave-function, and they severely degrade the signal-to-noise ratio because they only increase the statistical error.
Within the QMC scheme \eqref{Ph_scheme}, the overlap $\scal{\Psi_G}{\Phi}$ is given by
\begin{equation}
   \scal{\Psi_G}{\Phi} \proptauinf \brkt{\Psi_T}{\exp\Bigl(-\tau\H\Bigr)}{\Phi} \eqtauinf \bbE [\Pi_\tau] \;,
\end{equation}
$\Pi_\tau$ being defined by \eqref{Ph_scheme_2} but with $\Pi_0=\scal{\Psi_T}{\Phi}$.
Controlling the phase problem requires to use an approximation that will constrain the random walk by a change $\Pi\to\widetilde{\Pi}$ of the weight of the realizations.
First, by taking the real part of the local energy in its evolution \eqref{Ph_scheme_2}, we ensure that the overlaps $\scal{\Psi_G}{\Phi}$ with the ground state and $\scal{\Psi_T}{\Phi}$ with the trial state have the same phase.
Now, we have to prevent the random walk from populating symmetrically the complex $\scal{\Psi_T}{\Phi}$-plane to avoid the mean overlap merging with the origin.
To this end, we apply the \emph{phaseless} approximation \cite{Zhang_Phaseless}, where the constrained weight $\widetilde{\Pi}$ is supposed to evolve according to the dephasing $\d\theta=\arg \scal{\Psi_T}{\Phi_{\tau+\d\tau}}/\scal{\Psi_T}{\Phi_\tau}$ of the overlap with the trial state as
\begin{equation} \label{constraint}
   \widetilde{\Pi}_{\tau+\d\tau} = \widetilde{\Pi}_\tau\exp\Bigl(-\d\tau \mathrm{Re}\moy{\H}_{\Psi_T,\!\Phi_{\tau}} \Bigr)  \max\bigl\lbrace 0\,;\cos(\d\theta) \bigr\rbrace .
\end{equation}
Therefore, the more the phase of $\scal{\Psi_T}{\Phi}$ changes during $\d\tau$, the more the weight of the associated walker is decreased, and when $|\d\theta|>\pi/2$, the walker is discarded.
In this way, the centroid of the population is maintained into the half-plane $\mathrm{Re}\scal{\Psi_T}{\Phi}>0$ even if some realizations having reduced weights are generated in the other half-plane.
This constraint thus offers a compromise between the need to control the phase problem and the conservation of the initial form for the probability distribution.
Finally, replacing $\Pi$ by $\widetilde{\Pi}$ in \eqref{Ph_scheme} allows one to reconstruct without phase problem an approximate ground state $\ket{\widetilde{\Psi}_G}$.
We emphasize that approximation \eqref{constraint} also removes the risk of an infinite variance on the error $\ket{\Psi_G} -\ket{\Phi}$, which would result from an accumulation of realizations close to $\scal{\Psi_T}{\Phi}=0$.

The phaseless QMC method described above has been applied to many-body problems in quantum chemistry \cite{Zhang_Phaseless}, with a simple mean-field wave-function for the trial state $\ket{\Psi_T}$.
For the nuclear shell model, we propose here an improved scheme to reconstruct the ground state in each angular-momentum channel $J, M$.
Starting from a Slater determinant $\ket{\Phi_0}$, a satisfying test wave-function $\ket{\Psi^{JM}_T}$ to initiate, guide, and constrain the dynamics can be obtained by mixing all the possible values of the spin projection $K$ in the intrinsic frame:
\begin{equation} \label{Ansatz_SEMF}
   \ket{\Psi^{JM}_T}= \sum_{K=-J}^J C^J_K \hat{P}^J_{MK} \ket{\Phi_0}.
\end{equation}
Here, the operators $\hat{P}^J_{MK}$ are weighted averages of rotations $\hat{U}_\Omega$ parametrized with Euler's angles $\Omega$ as \cite{Villar}
\begin{equation} \label{Proj_PJKM}
   \hat{P}^J_{MK} = \dfrac{2J+1}{8\pi^2} \int\!\!\d\Omega\,D^{J*}_{MK}(\Omega)\hat{U}_\Omega ,
\end{equation}
where $D^{J}_{MK}$ denotes Wigner's $D$-function.
Below, we restrict the Slater determinant $\ket{\Phi_0}$ to be factorized into a product of independent-neutrons and -protons wave-functions so that the ansatz \eqref{Ansatz_SEMF} has also a good isospin projection.
Moreover, when the valence space contains only one major shell of the harmonic oscillator, the parity is already a good quantum number and no further restoration is needed.
Rotations $\hat{U}_\Omega$ simply transform $\ket{\Phi_0}$ into another Slater determinant.
As a superposition of symmetry-related Slater determinants, $\ket{\Psi^{JM}_T}$ is no longer an independent-particle state and can absorb correlations between particles beyond the mean-field level.
The energy $E_G^J$ of an yrast state $\ket{\Psi_G^{JM}}$ for the angular momentum $J$ can now be determined after a large enough imaginary-time as
\begin{equation} \label{yrast_Eg}
   E_G^J \!\eqtauinf \!\! \dfrac{\brkt{\Psi_T^{JM}}{\H\e^{-\tau\H}}{\Psi_T^{JM}}}{\brkt{\Psi_T^{JM}}{\e^{-\tau\H}}{\Psi_T^{JM}}}
    \!\! \eqtauinf \!\! \dfrac{\brkt{\Psi_T^J}{\H\e^{-\tau\H}}{\Phi_0}}{\brkt{\Psi_T^J}{\e^{-\tau\H}}{\Phi_0}} ,
\end{equation}
provided the approximate state $\ket{\Psi_T^{JM}}$ and the exact wave-function have a non-zero overlap.
In Eq. \eqref{yrast_Eg}, we have noted $\ket{\Psi_T^J} = \sum_{K,K'} C^{J*}_K C^J_{K'} \hat{P}^J_{KK'} \ket{\Phi}$ the new state that emerges from rotational invariance and from the well-known properties $(\hat{P}^J_{MK})^\dagger=\hat{P}^J_{KM}$ and $\hat{P}^J_{MK}\hat{P}^{J'}_{K'M'} = \delta_{JJ'}\delta_{KK'}\hat{P}^J_{MM'}$.
This wave-function appears as a natural candidate to apply the QMC formalism \eqref{Ph_scheme}-\eqref{Ph_scheme_2} with the biased weight \eqref{constraint} to constrain stochastic paths.
An approximate yrast state $\ket{\widetilde{\Psi}_G^{JM}}$ for any spin $J$ is thus obtained with energy $\widetilde{E}_G^J$ deduced from \eqref{Ph_scheme} and \eqref{yrast_Eg}
\begin{equation}
   \widetilde{E}_G^J \eqtauinf \dfrac{\bbE\Bigl[\widetilde{\Pi}_\tau \moy{\H}_{\Psi_T^J,\!\Phi_\tau}\Bigr]}{\bbE\bigl[\widetilde{\Pi}_\tau \bigr]}.  
\end{equation}
Note that a similar form also holds for any scalar observables commuting with the Hamiltonian.
In other cases, the well-known mixed estimate $\moy{\hat{A}}_{\Psi_T^{JM},\!\widetilde{\Psi}_G^{JM}}^{(\mathrm{mix})}$ gives an approximate ground-state expectation value of an observable $\hat{A}$:
\begin{equation}
   \moy{\hat{A}}_{\Psi_T^{JM},\!\widetilde{\Psi}_G^{JM}}^{(\mathrm{mix})} = \dfrac{\mathrm{Re}\brkt{\Psi_T^{JM}}{\hat{A}}{\widetilde{\Psi}_G^{JM}}} {\mathrm{Re}\scal{\Psi_T^{JM}}{\widetilde{\Psi}_G^{JM}}} .
\end{equation}
%
%
It can be corrected by the extrapolated estimator $\moy{\hat{A}}_{\widetilde{\Psi}_G^{JM}}^{(\mathrm{ext})} = 2 \moy{\hat{A}}_{\Psi_T^{JM},\!\widetilde{\Psi}_G^{JM}}^{(\mathrm{mix})} -   \moy{\hat{A}}_{\Psi_T^{JM}}$ that is one order of magnitude better in the difference $\ket{\widetilde{\Psi}_G^{JM}} - \ket{\Psi_T^{JM}}$.

In zero-temperature QMC methods, the imaginary time needed to converge to the ground-state wave-function depends on the quality of the approximate state used to initiate the propagation.
In the phaseless QMC scheme, this state $\ket{\Psi_T^{JM}}$ also guides and constrains the motion of walkers. All these considerations naturally lead to the choice of the variational solution obtained by energy minimization in the subspace of Slater determinants after quantum number projections.
Such a symmetry-entangled mean-field (SEMF) method was recently applied to the two-dimensional Hubbard model in condensed matter physics \cite{Juillet_SEMF} and in quantum chemistry \cite{Jimenez_PHF}.
In nuclear physics, the SEMF method is similar to the so-called VAMPIR approach \cite{Schmid_VAMPIR_SM}, but with a Slater determinant instead of a quasi-particle vacuum, and though the energy minimization is carried out in a different way.
First, we note that the energy $E_T^J$ in the trial state \eqref{Ansatz_SEMF} reads
\begin{equation} \label{E_SEMF}
   E_T^J =\dfrac{\sum_{K,K'} C^{J*}_K C^J_{K'} \brkt{\Phi_0}{\H\hat{P}^J_{KK'}}{\Phi_0}}{\sum_{K,K'} C^{J*}_K C^J_{K'} \brkt{\Phi_0}{\hat{P}^J_{KK'}}{\Phi_0}},
\end{equation}
so that variation with respect to the amplitudes $C^{J*}_K$ yields a generalized eigenvalue equation:
\begin{equation} \label{Val_P_CK}
   \sum_{K'} C^J_{K'}\brkt{\Phi_0}{\H\hat{P}^J_{KK'}}{\Phi_0} = E_T^J \sum_{K'} C^J_{K'} \brkt{\Phi_0}{\hat{P}^J_{KK'}}{\Phi_0} .
\end{equation}
On the other hand, owing to Wick's theorem \cite{Blaizot_Ripka}, the expectation values of $\H\hat{P}^J_{KK'}$ and $\hat{P}^J_{KK'}$ in $\ket{\Phi_0}$ are only expressed with the contractions $(\rho_0)_{ij}=\moy{\cc{j}\ca{i}}_{\Phi_0}$, $\rho_0$ being the one-body density matrix associated with $\ket{\Phi_0}$.
Therefore, the energy \eqref{E_SEMF} is a functional of $\rho_0$, i.e. $E_T^J=E_T^J[\rho_0]$.
The stationarity condition under the constraint of orthonormal single-particle states $\{\ket{\phi_{0,n}}\}$ immediately leads to the following Hartree-Fock like equation
\begin{equation} \label{HF_eq}
   \Bigl[ h^J[\rho_0]\,,\rho_0\Bigr]=0 .
\end{equation}
The effective single-particle SEMF Hamiltonian $h^J$ is defined as the gradient of the projected energy according to the density, $h^J_{ij} = \partial E_T^J/\partial(\rho_0)_{ji}$.
This derivative can be further calculated by using Wick's theorem extended for matrix elements \cite{Blaizot_Ripka} through the introduction of one-body density matrices $\R_\Omega$ between $\ket{\Phi_0}$ and $\hat{U}_\Omega\ket{\Phi_0}$.
The SEMF Hamiltonian is finally given by:
\begin{equation} \label{hSEMH}
  \begin{aligned}
     &  h^J[\rho_0] = \int\!\!\d\Omega X^J_\Omega A_\Omega^{-1} \left[ \bigl(U_\Omega-1\bigr) \bigl(\mathcal{E}[\R_\Omega] - E^J \bigr) \right.\\ 
     &  \qquad\qquad\qquad\qquad\quad\left.\quad + \: h[\R_\Omega]U_\Omega B_\Omega^{-1} \right], \\
     & X^J_\Omega = \dfrac{\sum_{KK'} C^{J*}_K C^J_{K'} D^{J*}_{KK'}(\Omega) \det(A_\Omega)}{\int\!\!\d\Omega \sum_{KK'} C^{J*}_K C^J_{K'} D^{J*}_{KK'}
(\Omega)\det(A_\Omega)}.
   \end{aligned}
\end{equation}
Here, $A_\Omega$ and $B_\Omega$ depends on $\rho_0$ and on the matrix representation $U_\Omega$ of rotations in the one-particle space according to $A_\Omega = 1 + (U_\Omega -1)\rho_0$ and $B_\Omega = 1 + \rho_0 (U_\Omega -1)$;
$\mathcal{E}[\R_\Omega]$ and $h[\R_\Omega]$ stands for the usual Hartree-Fock energy functional and Hamiltonian, but in which the density is replaced by $\R_\Omega$.
This transition density matrix can be expressed as $\R_\Omega=\rho_0 U_\Omega A_\Omega^{-1}$.
More details about the derivation of \eqref{hSEMH} can be found in Refs. \cite{Jimenez_PHF,Sheikh_PHFB}.

\begin{figure}[!t]
\begin{center}
\includegraphics[scale=0.375]{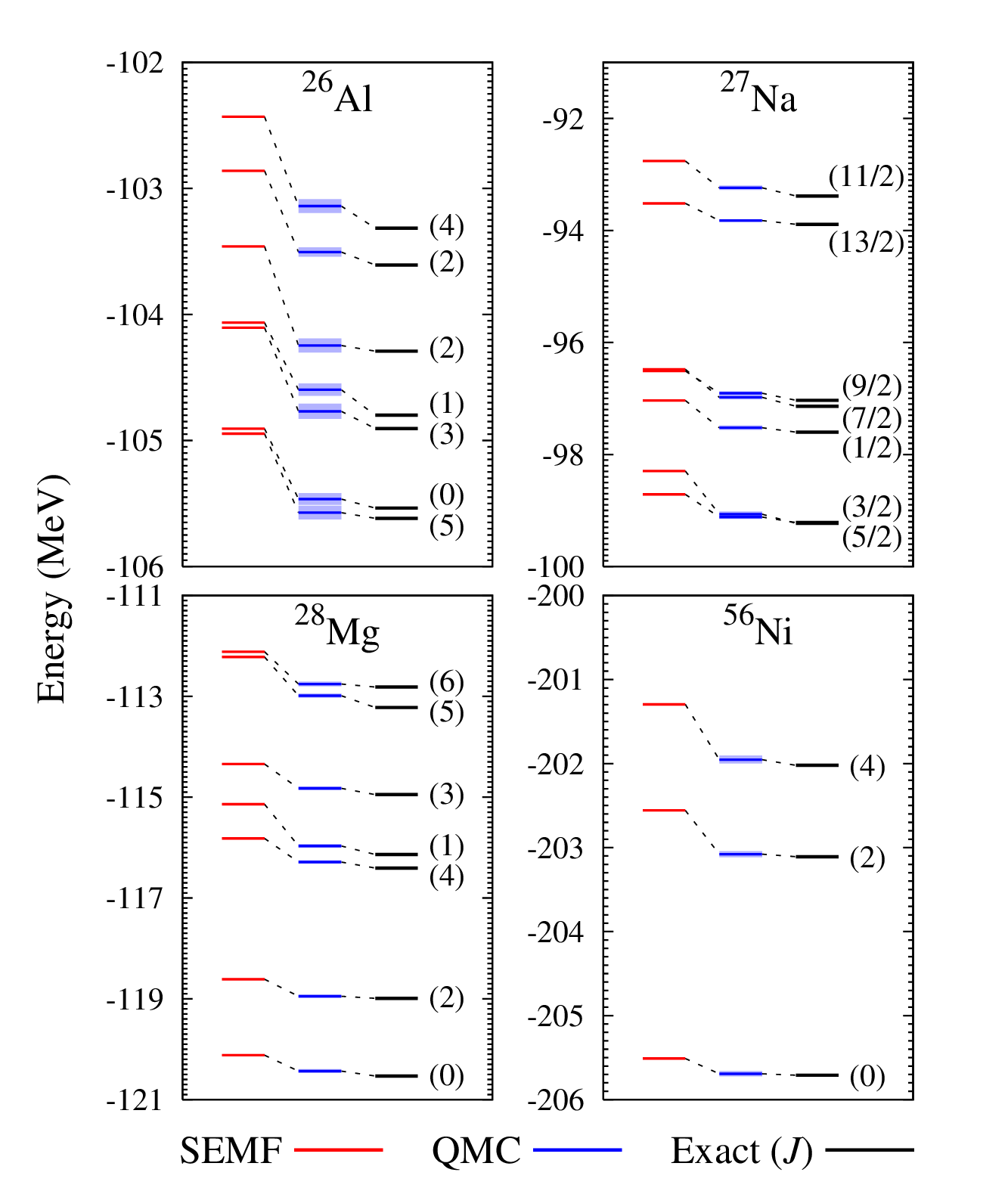}
\caption{\textbf{(Color online)} Yrast spectra of three $sd$-shell nuclei and one $pf$-shell nucleus as obtained with the SEMF and QMC methods compared to the exact energy levels.
The lighter areas indicate the QMC statistical errors.}
\label{fig_spectra}
\end{center}
\begin{center}
\subfigure{
\includegraphics[scale=0.166,angle=-90]{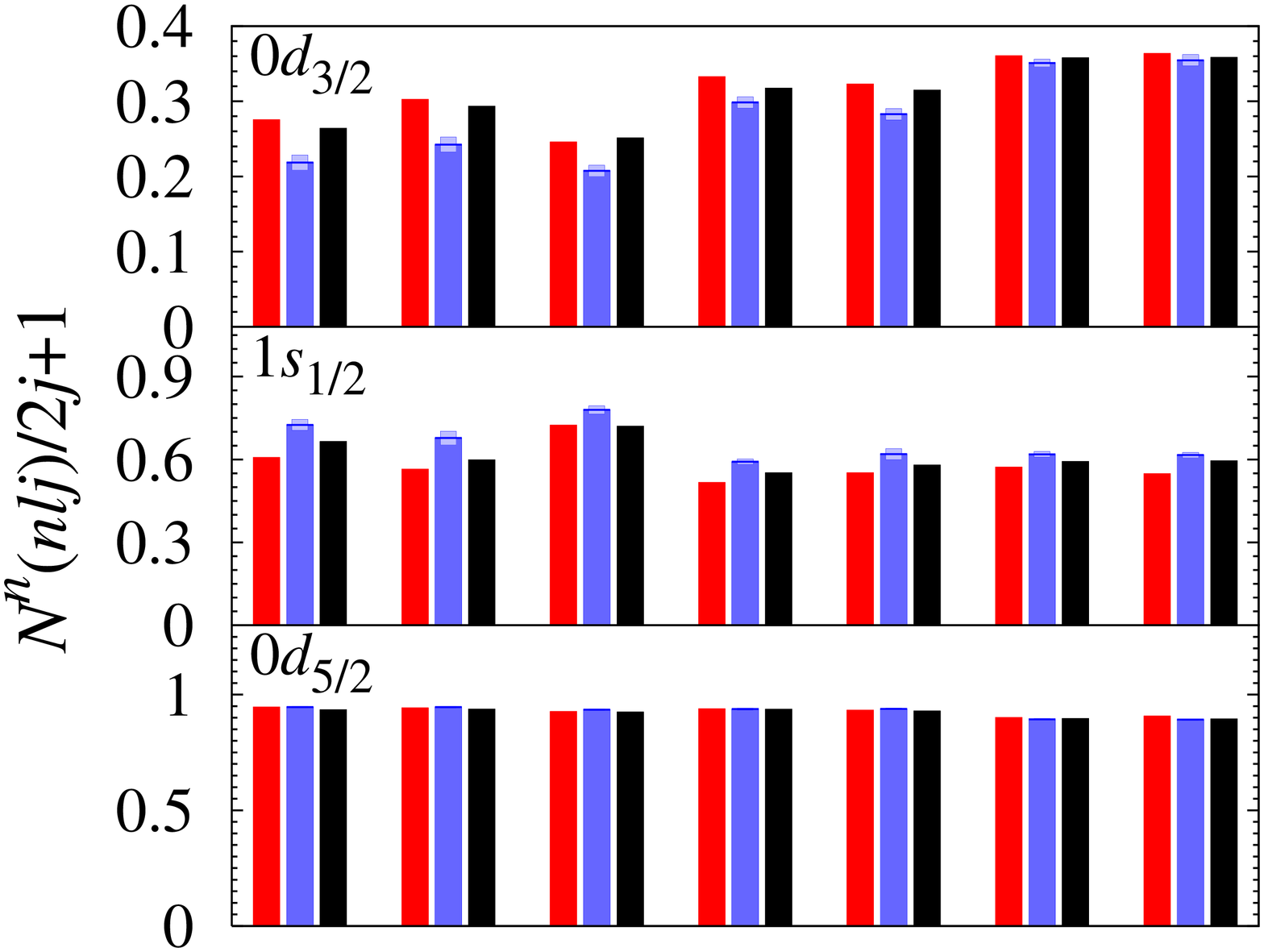}
\includegraphics[scale=0.166,angle=-90]{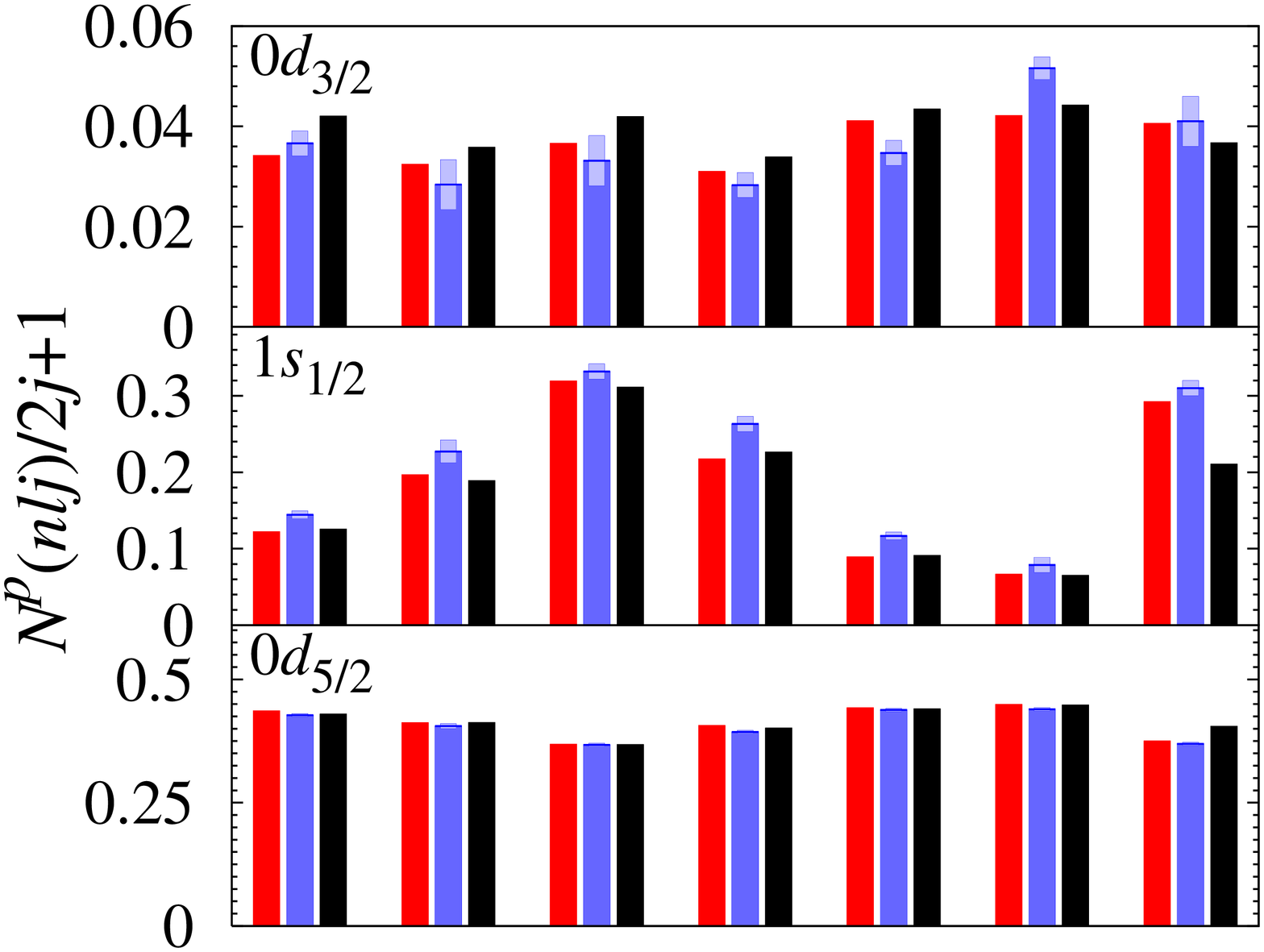}}
\subfigure{
\includegraphics[scale=0.166,angle=-90]{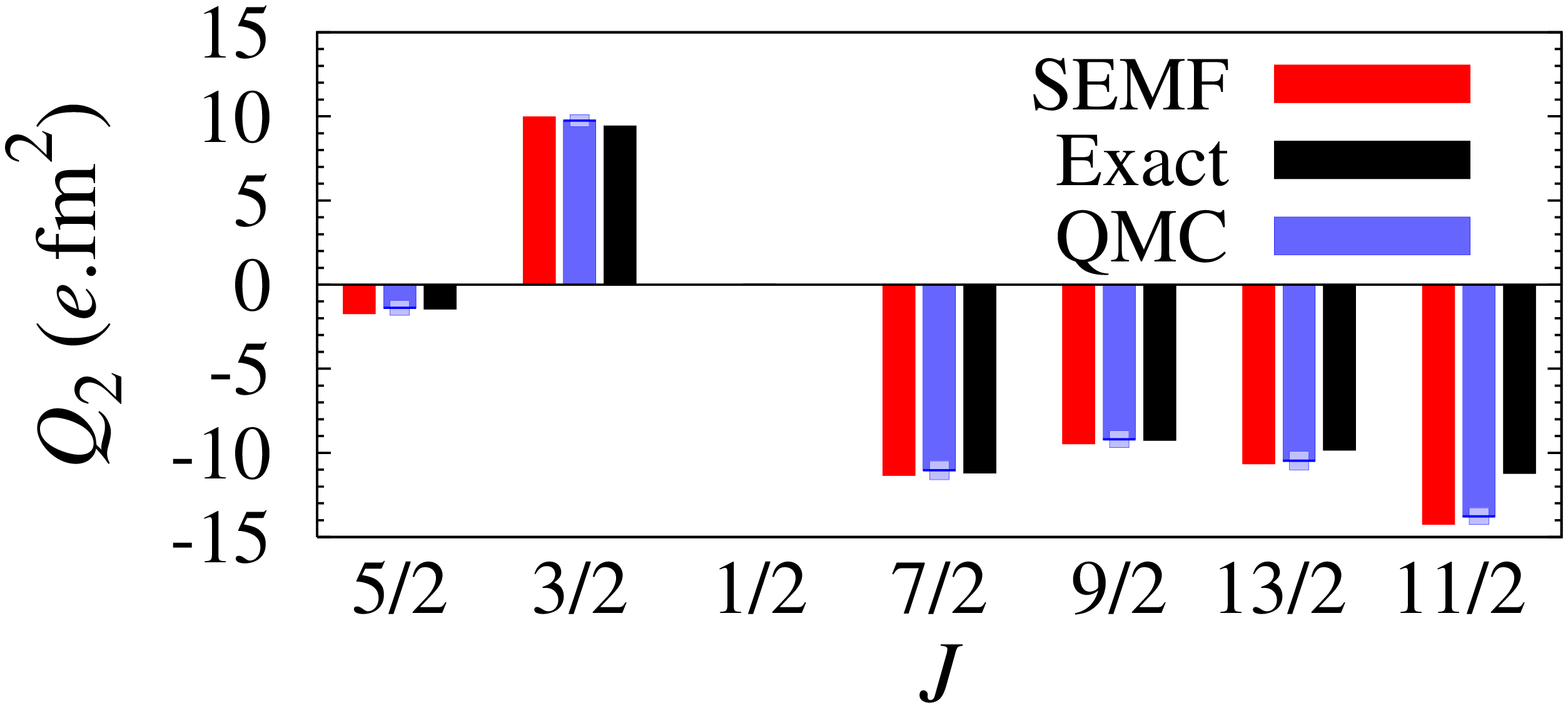}
\includegraphics[scale=0.166,angle=-90]{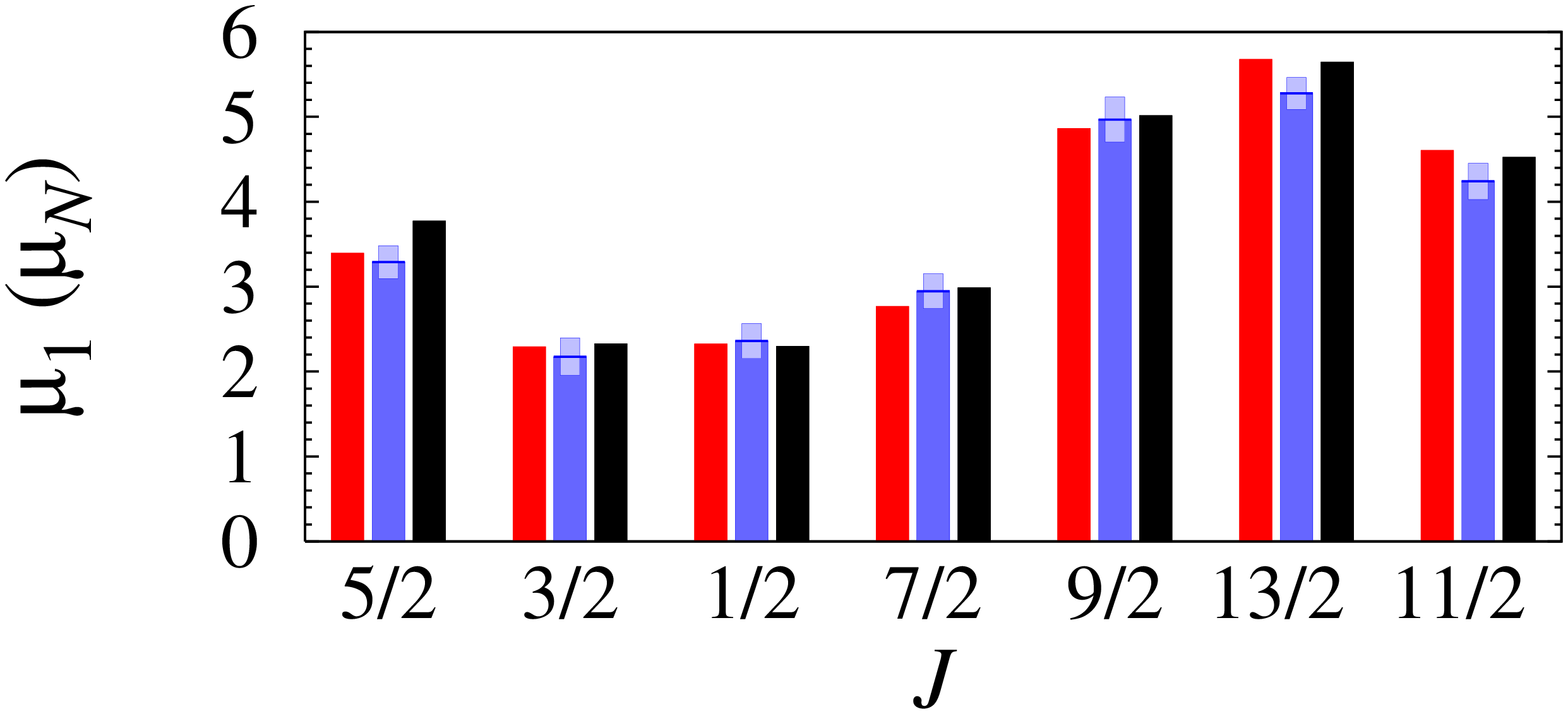}}
\caption{\textbf{(Color online)} SEMF, QMC (extrapolated estimate), and exact values of some observables for the yrast states of ${}^{27}$Na of Fig.~\ref{fig_spectra}: Neutron (upper left) and proton (upper right) occupations of the shells $nlj$ normalized at $2j+1$, electric quadrupole moment (lower left), and magnetic dipole moment.
The effective charges, orbital and spin $g$-factors have been chosen to be respectively $e_n=0.49e$, $g_n^l=0.036$, $g_n^s=-3.875$ for neutrons, and $e_p=1.29e$, $g_p^l=1.03$, $g_p^s=5.586$ for protons.}
\label{fig_obs_27Na}
\end{center}
\end{figure}

Let us now present phaseless QMC results for the nuclear shell model.
We address here systems for which exact results are available for benchmark.
Fig.~\ref{fig_spectra} displays the SEMF, QMC, and exact yrast spectra of three kinds of $sd$-shell nuclei with the USD interaction \cite{Wildenthal_USD}: A $N=Z$ odd-odd nucleus (${}^{26}$Al), an odd-mass nucleus (${}^{27}$Na), and an even-even nucleus (${}^{28}$Mg).
Also shown is the spectrum of ${}^{56}$Ni in the $pf$ model space with the GXPF1A interaction \cite{Honma_GXPF1A}.
Exact energies are extracted from \cite{Pittel_56Ni}.
First, we observe in Fig.~\ref{fig_spectra} that the SEMF method offers a good approximation in all considered cases: Yrast energies differ from the exact values by less than about 1 MeV.
Second, relative errors on the QMC energies at any angular momentum do not exceed 0.3 {\%} with statistical error bars about 40-50 keV.
In particular, the results for ${}^{27}$Na give good evidence that the phase problem is well controlled in our calculations. 
Indeed, the odd-mass nuclei are  more pathological with respect to QMC simulations, because of the time-reversal symmetry breaking of the stochastic dynamics that leads to a phase problem irrespective of the interaction.

To further test the nuclear phaseless QMC scheme, we now examine the neutron (proton) occupation numbers $N^{n(p)}$ of the various shells ($nlj$), the electric quadrupole $Q_2$ and magnetic dipole $\mu_1$ moments.
The results for these observables as obtained with the SEMF and QMC methods are compiled in Fig.~\ref{fig_obs_27Na} for the nucleus ${}^{27}$Na and compared to exact diagonalization.
As above, the SEMF approach again provides a good approximation, the agreement with the exact values being roughly correct for any angular momentum.
Nevertheless, in contrast to the case of the energy, no significant improvement of the SEMF results is observed after the imaginary-time propagation for these observables that do not commute with the Hamiltonian.
This may point out the necessity to go beyond extrapolated estimates by using, for instance, the back-propagation techniques \cite{Purwanto_BP}.
Finally, for all the states of Fig.~\ref{fig_spectra}, we have found exact expectation values of the squared isospin.

In summary, we have presented a new QMC formalism for the nuclear shell model. 
The method relies on a mean-field wave-function entangled by symmetry restoration to initiate, guide, and also constrain the Brownian motion in order to control sign/phase problems.
The results reported in this Letter demonstrate that the phaseless QMC approach can accurately reproduce the yrast spectroscopy of $sd$- and $pf$-shell nuclei with realistic effective interactions.
Electromagnetic transitions and $\beta$ decay can also be considered \textit{via} a specific mixed-estimator \cite{Pervin_MixTrans}.
Finally, calculations on medium-mass neutron-rich and proton-rich nuclei are under development to investigate low-lying structure in the full $pf-0g_{9/2}$ shell that can only be addressed by conventional diagonalization methods through severe truncations of allowed configurations.
We are also extending the phaseless QMC approach for computing excited states of a given angular momentum in order to achieve a complete spectroscopy of nuclei.

We gratefully thank R. Fr{\'e}sard and P. van Isacker for a careful reading of the manuscript.



\begin{thebibliography}{99}

\bibitem{Caurier_SM} E. Caurier, G. Martínez-Pinedo, F. Nowacki, A. Poves, and A. P. Zuker, Rev. Mod. Phys. \textbf{77}, 427 (2005).

\bibitem{Koonin_SMMC} S. E. Koonin, D. J. Dean, and K. Langanke, Phys. Rep. \textbf{278}, 1 (1999), and references therein.
\bibitem{Ozen_SMMC} C. {\"O}zen, Y. Alhassid, and H. Nakada, Phys. Rev. Lett. \textbf{110}, 042502 (2013).
\bibitem{Alhassid_SMMC_SP} Y. Alhassid, D. J. Dean, S. E. Koonin, G. Lang, and W. E. Ormand, Phys. Rev. Lett. \textbf{72}, 613 (1994).


\bibitem{Honma_MCSM} M. Honma, T. Mizusaki, and and T. Otsuka, Phys. Rev. Lett. \textbf{77}, 3315 (1996).
\bibitem{Otsuka_MCSM} T. Otsuka, T. Mizusaki, N. Shimizu, and Y. Utsuno, Prog. Part. Nucl. Phys. \textbf{47}, 319 (2001).

\bibitem{Pieper_GFMC} S. C. Pieper and R. B. Wiringa, Ann. Rev. Nucl. Part. Sci. \textbf{51}, 53 (2001).

\bibitem{Zhang_Phaseless} S. Zhang and H. Krakauer, Phys. Rev. Lett. \textbf{90}, 136401 (2003).

\bibitem{Gardiner} C. W. Gardiner, \textit{Handbook of Stochastic Methods} (Springer-Verlag, Berlin, 1983).

\bibitem{Villar} F. Villar, \textit{Varenna Lectures Vol. 36} (Bloch, Academic, New York, 1966).

\bibitem{Juillet_SEMF} O. Juillet and R. Fr{\'e}sard, Phys. Rev. B \textbf{87}, 115136 (2013).

\bibitem{Jimenez_PHF} C. A. Jim{\`e}nez-Hoyos, T. M. Henderson, T. Tsuchimochi, and G. E. Scuseria, J. Chem. Phys. \textbf{136}, 164109 (2012).

\bibitem{Schmid_VAMPIR_SM} K. W. Schmid, Prog. Part. Nucl. Phys. \textbf{52}, 565, and references therein.

\bibitem{Blaizot_Ripka} J. P. Blaizot and G. Ripka, \textit{Quantum Theory of Finite Systems}, (MIT Press, Cambridge, Massachusetts, 1986).

\bibitem{Sheikh_PHFB} J. A. Sheikh, P. Ring, E. Lopes, and R. Rossignoli, Phys. Rev. C \textbf{66}, 044318 (2002).

\bibitem{Wildenthal_USD} B. H. Wildenthal,  Prog. Part. Nucl. Phys. \textbf{11}, 5 (1984).

\bibitem{Honma_GXPF1A} M. Honma, T. Otsuka, B. A. Brown, and T. Mizusaki, Eur. Phys. J. A. \textbf{25}, 499 (2005).

\bibitem{Pittel_56Ni} S. Pittel and B. Thakur, Rev. Mex. F{\'i}s. \textbf{55}, 108 (2009).

\bibitem{Purwanto_BP} W. Purwanto and S. Zhang, Phys. Rev. E \textbf{70}, 056702 (2004).

\bibitem{Pervin_MixTrans}  M. Pervin, S. C. Pieper, and R. B. Wiringa, Phys. Rev. C \textbf{64}, 014001 (2001).
\end{thebibliography}
\end{document}